# Dry release transfer of graphene and few-layer h-BN by utilizing thermoplasticity of polypropylene carbonate for fabricating edge-contact-free van der Waals heterostructures

Kei Kinoshita[1], Rai Moriya[1,*], Momoko Onodera[1], Yusai Wakafuji[1], Satoru Masubuchi[1], Kenji Watanabe[2], Takashi Taniguchi[2], and Tomoki Machida[1,*]

[1] *Institute of Industrial Science, University of Tokyo, 4-6-1 Komaba, Meguro, Tokyo 153-8505, Japan*

[2] *National Institute for Materials Science, 1-1 Namiki, Tsukuba 305-0044, Japan*

The dry release transfer of two-dimensional (2D) materials such as graphene, hexagonal boron nitride (h-BN), and transition metal dichalcogenides (TMDs) is a versatile method for fabricating high-quality van der Waals heterostructures. Up until now, polydimethylpolysiloxane (PDMS) sheets have been widely used for the dry release transfer of TMD materials. However, this method has been known to have limitations that make it difficult to transfer few-layer-thick graphene and h-BN because of the difficulty to fabricate these materials on PDMS. As an alternative method, we demonstrate the dry release transfer of single- and bi-layer graphene and few-layer h-BN in this study by utilizing poly(propylene) carbonate (PPC) films. Because of the strong adhesion between PPC and 2D materials around room temperature, we demonstrate that single- to few-layer graphene, as well as few-layer h-BN, can be fabricated on a spin-coated PPC film/290-nm-thick SiO$_2$/Si substrate via the mechanical exfoliation method. In addition, we show that these few-layer **crystals are clearly distinguishable using an optical microscope with the help of optical interference. Because of the thermoplastic properties of PPC film, the adhesion force between the 2D materials and PPC significantly decreases at about 70 °C. Therefore, we demonstrate that single- to few-layer graphene, as well as few-layer h-BN flakes, on PPC can be easily dry-transferred onto another h-BN substrate. This method enables a multilayer van der Waals heterostructure to be constructed with a minimum amount of polymer contamination. We demonstrate the fabrication of encapsulated h-BN/graphene/h-BN devices and graphene/few-layer h-BN/graphene vertical-tunnel-junction devices using this method. Since devices fabricated by this method do not require an edge-contact scheme, our finding provide a simples method for constructing high-quality graphene and h-BN-based van der Waals heterostructures.**

*E-mail: moriyar@iis.u-tokyo.ac.jp; tmachida@iis.u-tokyo.ac.jp

# 1. Introduction

Up until now, various kinds of van der Waals (vdW) heterostructures composed of graphene, hexagonal boron nitridee (h-BN), and other two-dimensional (2D) materials have been fabricated for the purposes of fundamental research and applications [1,2]. To construct these heterostructures, various transfer methods for 2D materials have been explored [3], including the (1) wet release transfer [4], (2) dry release transfer [5-7], and (3) stamping methods [8-11]. Among them, the dry release transfer method has been recognized as the most versatile and flexible technique to explore various kinds of heterostructures. In this method, flakes of the 2D material are first fabricated on a polymer sheet and are subsequently transferred onto another material (typically, exfoliated h-BN on a Si substrate) without dissolving the polymer sheets by organic solvents. The most commonly used polymer for the dry transfer method is polydimethylpolysiloxane (PDMS) [5]. Various vdW heterostructures have been fabricated based on this method for utilizing transition metal dichalcogenides such as $MoS_2$, $WSe_2$, and $NbSe_2$, as well as black phosphorus. Examples of the applications of these heterostructures are high-quality quantum Hall effect devices [12,13], superconducting devices [14-17], optoelectronic devices [18-21], and spintronics devices [22-24]. However, the PDMS-based dry transfer method has a limited capability for use in single, bi-layer, and few-layer-thick graphene and few-layer-thick h-BN because of the poor adhesion between PDMS and these materials which makes it difficult to fabricate these thin crystals on PDMS sheets, and owing to the poor visibility of the thin flakes prepared on PDMS sheet [5,25-27]. Alternative polymers that enable reliable exfoliation and dry release transfer of thin graphene and h-BN have been in high demand. In this study, we demonstrate the dry release transfer method using

poly(propylene) carbonate (PPC) for fabricating graphene- and h-BN-based vdW heterostructures. The main advantages of this method are summarized in the following three points. 1) Because of the strong adhesion of PPC to graphene and h-BN at room temperature (RT), appropriately large flakes of thin graphene and h-BN can be obtained with the mechanical exfoliation method. 2) Single-layer graphene, as well as thin h-BN, fabricated on a PPC/SiO$_2$/Si structure is visible under an optical microscope. 3) As the adhesion between PPC and the flakes significantly decreases at high temperatures, we demonstrate the dry release transfer of graphene and h-BN on a h-BN/SiO$_2$/Si substrate at about 70 °C. The proposed dry release transfer method is suitable for fabricating graphene- and h-BN-based vdW heterostructures.

## 2. Calculation of optical contrast

First, we show the numerical calculation of optical contrast for graphene and h-BN on different substrates, and the results are presented in Fig. 1. The optical contrast for each substrate was calculated based on the multilayer Fresnel reflection method [28-30]. We consider the case of visible light incidence from air onto the multi-layer structure including graphene, h-BN, PPC, SiO$_2$, and Si. The refractive index values of graphene, PPC, SiO$_2$, and Si are taken from the literature [31-33]. We consider the results for graphene/PPC/SiO$_2$/Si as depicted in Fig. 1(a). The single-layer graphene is assumed to have a thickness $d_1$ = 0.34 nm and complex wavelength-independent refractive index $n_1(\lambda)$ = 2.6–1.3i [28]. For multilayer graphene, we use the thickness of $Nd_1$, where $N$ represents the number of layers. We changed the thickness of the PPC layer $d_2$ from 0 to 1000 nm with a refractive index of $n_2(\lambda)$ = 1.463. The thickness of the SiO$_2$ layer $d_3$ is fixed at $d_3$ =

290 nm with a wavelength-dependent refractive index $n_3(\lambda)$ consisting of only the real part only (typically, $n_3(\lambda) \sim 1.458$ for $\lambda \sim 600$ nm). The thickness of the Si layer is assumed to be semi-infinite with a wavelength-dependent complex refractive index $n_4(\lambda)$ (the refractive index data for Si and SiO$_2$ are provided in the supplementary information). Using the geometry depicted in Fig. 1(a), the reflected light intensity can be written as

$$I(n_1) = \left| \frac{e^{i\delta_3}\{e^{i\delta_2}(r_0 e^{i\delta_1}+r_1 e^{-i\delta_1})+r_2 e^{-i\delta_2}(e^{-i\delta_1}+r_0 r_1 e^{i\delta_1})\}+r_3 e^{-i\delta_3}\{e^{-i\delta_2}(e^{-i\delta_1}+r_0 r_1 e^{i\delta_1})+r_2 e^{i\delta_2}(r_0 e^{i\delta_1}+r_1 e^{-i\delta_1})\}}{e^{i\delta_3}\{e^{i\delta_2}(e^{i\delta_1}+r_0 r_1 e^{-i\delta_1})+r_2 e^{-i\delta_2}(r_0 e^{-i\delta_1}+r_1 e^{i\delta_1})\}+r_3 e^{-i\delta_3}\{e^{-i\delta_2}(r_0 e^{-i\delta_1}+r_1 e^{i\delta_1})+r_2 e^{i\delta_2}(e^{i\delta_1}+r_0 r_1 e^{-i\delta_1})\}} \right|^2,$$

where

$$r_0 = \frac{n_0 - n_1}{n_0 + n_1}$$

$$r_1 = \frac{n_1 - n_2}{n_1 + n_2}$$

$$r_2 = \frac{n_2 - n_3}{n_2 + n_3}$$

are the reflection coefficients for different interfaces, and $\delta_1 = 2\pi n_1 d_1/\lambda$, $\delta_2 = 2\pi n_2 d_2/\lambda$, and $\delta_3 = 2\pi n_3 d_3/\lambda$ characterize the phase shifts when light passes through the $n$th layer. Then, the optical contrast $C$ is defined as the relative intensity of reflected light in the presence ($n_1 \neq 1$) and absence ($n_1 = n_0 = 1$) of graphene:

$$C = \frac{I(n_1 = 1) - I(n_1)}{I(n_1 = 1)}.$$

The results are plotted as a function of the incident light wavelength and PPC thickness, as shown in Fig. 1(b). The contrast curve for single and bi-layer graphene on PPC with a thickness of 900 nm is presented in Fig. 1(c). Contrast exhibits zero to positive values and it depends on wavelength. The graphene-covered region exhibits significant absorption for green, blue, and red light, thus rendering a region on the image that is darker than the surrounding area. This demonstrates that graphene on the PPC/SiO$_2$/Si substrate can be distinguishable with an optical microscope. For comparison, the optical contrast for graphene/SiO$_2$/Si, as depicted in Fig. 1(f), is calculated. The variation in optical contrast for different thicknesses of SiO$_2$ and wavelengths is plotted in Fig. 1(g), and the results for single- and bi-layer graphene on 290-nm-thick SiO$_2$/Si are presented in Fig. 1(h). The calculated optical contrast in Fig. 1(h) is fully consistent with the literature and clear layer-dependent contrast is exhibited [28,29]. The contrast curves in Figs. 1(b) and 1(g) are quite similar. This is due to the similarity of the refractive index of PPC to that of SiO$_2$; thus, the spin-coated PPC on SiO$_2$/Si gives rise to a similar effect with the increase in SiO$_2$ thickness in the SiO$_2$/Si structure. Therefore, it is a natural consequence that few-layer graphene is visible on a PPC/SiO$_2$/Si substrate with a moderately thin PPC, as studied here. Optical micrographs for single- and bi-layer graphene on 900-nm-thick PPC/290-nm-thick SiO$_2$/Si are shown in Figs. 1(d) and 1(e), respectively. We find that single- and bi-layer graphene on PPC/SiO$_2$/Si can be visible under an optical microscope. For comparison, optical micrographs of single- and bi-layer graphene on 290-nm-thick SiO$_2$/Si taken with the same optical microscope are also shown in Figs. 1(i) and 1(j), respectively. Here, images were taken in high-dynamic-range mode to enhance the visibility of the micrograph.

We also calculated the contrast for the h-BN/PPC/290-nm-thick SiO$_2$/Si structure, as illustrated in Fig. 1(k). Here, the thickness for h-BN is assumed to be $d_5$ = 0.335 nm with a wavelength-independent refractive index of $n_5(\lambda)$ = 2.2 [34]. The results are plotted as a function of the incident light wavelength and PPC thickness, as shown in Fig. 1(l). The contrast curves for single and bi-layer h-BN on PPC with a thickness of 900 nm are presented in Fig. 1(m). For comparison, the optical contrast for h-BN/SiO$_2$/Si, as depicted in Fig. 1(p), is calculated. The variation in optical contrast for different thicknesses of SiO$_2$ and wavelengths is plotted in Fig. 1(q), and the results for single- and bi-layer h-BN on 290-nm-thick $SiO_2$/Si are presented in Fig. 1(r). For both the PPC/$SiO_2$/Si and $SiO_2$/Si substrate, h-BN exhibits a positive or negative optical contrast depending on the wavelength. These positive and negative contrasts could easily cancel out each other under optical microscope observation with white light. Thus, compared to graphene, few-layer h-BN is more difficult to distinguish with a microscope. Previous experiments on few-layer h-BN on a $SiO_2$/Si substrate used optical filters and optimization of the $SiO_2$ thickness to enhance visibility [34]. Because of the similarity between Figs. 1(l) and 1(q), we believe that few-layer h-BN can be easily distinguishable using filters and by optimizing the thickness of both PPC and $SiO_2$. Nevertheless, we show that even without optimization, the technologically important three- and four-layer h-BN can be distinguishable on a 900-nm-thick PPC/$SiO_2$/Si substrate, and their optical micrographs are presented in Figs. 1(n) and 1(o). These results indicate that the PPC/$SiO_2$/Si structure enables atomically thin 2D crystals to be exfoliated, and their thickness can be identified with an optical microscope.

**3. Dry transfer process**

Here, we show a method to demonstrate the dry transfer of single-layer graphene on a h-BN substrate. As the thin graphene and h-BN are visible on the PPC film, we used this 2D material/PPC stack for the dry transfer of 2D materials onto another substrate. The detailed transfer procedure for fabricating the graphene/h-BN structure is illustrated in the schematic in Fig. 2. First, the PPC solution was spin-coated on the $SiO_2$/Si substrate at a spin-coating speed of 4000 rpm; this created a PPC film with a thickness of about 900 nm. Prior to spin coating, the substrate was cleaned by acetone and IPA with ultrasonic agitation.

Then, the PPC-coated substrate was baked on a hot plate at 70 °C for 5 min. The PPC/SiO$_2$/Si substrate was taken away from the hot plate and graphene or h-BN was then deposited on the substrate by the mechanical exfoliation method at RT, as shown in Fig. 2(a). As PPC is known to exhibit a strong adhesion to 2D materials at around RT [8,10], both few-layer graphene and thick graphite or h-BN with a reasonably large size can be easily fabricated. Separately, a piece of PDMS sheet with a size of approximately 3 mm (width) × 3 mm (length) × 0.4 mm (depth) was placed on a glass slide. To ensure a strong adhesion between the PDMS and the glass slide, the backside of the PDMS was treated with air plasma for a few minutes. On the PDMS sheet, graphene flakes on the PPC film were transferred from the SiO$_2$/Si substrate by the following procedure, as depicted in Figs. 2(b) and 2(c): 1) prepare the tape window with a size of about 4 mm × 4 mm on the PPC that surrounds the graphene region; 2) clean the front side of the PDMS sheet with air plasma for a few minutes to ensure a strong adhesion of the PDMS surface; 3) remove the graphene/PPC membrane structures together with the tape by gently removing the tape from the substrate; 4) attach the graphene/PPC membrane on the PDMS sheet by aligning the tape window to the PDMS sheet (Fig. 2(c)). Separately, h-BN flakes were prepared on another SiO$_2$/Si substrate by mechanical exfoliation (Fig. 2(d)). The substrate was pre-cleaned with piranha solution prior to the deposition of h-BN. For transfer, the relative positions of graphene/PPC and h-BN were aligned under optical microscope observation, as shown in Fig. 2(e), and the graphene and h-BN were made to gently come into contact each other without heating up the stage. Once they made contact, the substrate stage was heated up to 70 °C (Fig. 2(f)). Here, we changed only the setpoint of the substrate heater from RT to 70 °C; then, the substrate heated up with the maximum speed of the heater.

When the substrate stage temperature was stable at 70 °C, the substrate and PPC film were slowly separated by moving either the substrate stage or the glass slide stage (Fig. 2(g)). We found that at temperatures equal to or higher than about 70 °C, the adhesion between graphene and PPC significantly weakens such that graphene can be released from the PPC film and transferred onto h-BN. Thus, the graphene/h-BN structure can be fabricated. Similarly, thick or thin h-BN can be transferred onto the h-BN or graphene. The dry transfer operates within the temperature range of 70–100 °C, depending on the adhesion between PPC and PDMS. The PPC film tends to detach from PDMS during transfer when the substrate temperature is higher than 100 °C. The process does not require graphene (or h-BN) to be exposed to organic solvents; thus, this can be regarded as a dry release transfer. To completely remove the PPC residue from the graphene surface, the sample was annealed at 350 °C in an Ar/3% $H_2$ atmosphere for 1 h; this temperature is above the decomposition temperature of PPC of about 280 °C [35]. The air-plasma treatment process during transfer can be replaced with an $O_2$ plasma or an equivalent process to improve the adhesion at the PPC/PDMS and PDMS/glass interfaces. We also note that these plasma treatment processes are not crucial steps to achieve flake transfer but they reduce the risk of PDMS or PPC sheet falling off from the glass slide or PDMS during transfer.

Here, we discuss the mechanism of the PPC-based dry transfer. PPC is known as a thermoplastic material such that it becomes very soft at higher temperatures, whereas it hardens at RT. To compare this to PDMS, we plot the storage modulus of these materials taken from the literature in Fig. 3(a) [31,36]. The storage modulus represents the mechanical hardness of the material. The glass transition temperature ($T_g$) of PPC is about 40 °C. Below this temperature, the storage modulus rapidly increases and saturates at a lower temperature. Here, PPC behaves as a solid and is difficult to deform. The adhesion between the 2D crystal and PPC is known to be strong in this region. In fact, the pick-up of thick h-BN from the SiO$_2$/Si substrate with a PPC sheet has been performed around this temperature regime [8,10]. We demonstrate that reasonably large graphene and h-BN flakes can be easily fabricated with mechanical exfoliation. An optical micrograph of graphene and graphite on PPC fabricated under this condition is presented in Figs. 3(b) and 3(c), respectively. Heating the PPC above about 40 °C drastically reduces its storage modulus. Here, PPC is glass and becomes soft. The adhesion between the 2D crystal and PPC gradually decreases with the increase in temperature above this region. At high temperatures (about 70 °C) where the adhesion between PPC and graphene or PPC and h-BN is sufficiently small, these crystals tend to detach from the PPC and transfer to another substrate. Here, graphene and the graphite flakes respectively shown in Fig. 3(b) and 3(c) are transferred onto the 290-nm-thick SiO$_2$/Si substrate using the method explained in Fig. 2, and their optical micrographs are presented in Figs. 3(d) and 3(e), respectively. Nearly all the graphene and graphite flakes can be transferred from PPC to the SiO$_2$ substrate. This indicates that our dry transfer method relies on the strong reduction of adhesion between the 2D crystal and PPC membrane at high temperatures. Thus, the flakes on the PPC can be easily transferred even on the SiO$_2$ surface. For comparison, the storage modulus curve for PDMS is also presented in Fig. 3(a). As the $T_g$ of PDMS is –28 °C, PDMS does not show a dramatic change in storage modulus with the increase in temperature and is always in the glass state and soft above RT. Because of this softness, PDMS is very effective for the dry transfer of the flake once the flake has been exfoliated onto the PDMS surface. The drawback of this method is the difficulty of the exfoliation step of the thin graphene and h-BN flakes onto PDMS due to the poor adhesion. The advantage of using the PPC presented here is the strong adhesion at RT, which becomes very weak at higher temperatures. This is in contrast to the PMMA-based dry release transfer method, where the transfer relies on the strong vdW interaction between the 2D crystals and h-BN substrate [4,6,7]. Therefore, dry transfer can be successful only where h-BN is present; the dry transfer of 2D crystals (such as graphene) onto $SiO_2$/Si is difficult with the PMMA-based process. In addition, the PMMA-based dry transfer requires a sacrificed polymer layer between PMMA and the substrate to peel the PMMA off from the substrate [6,7]; a sacrificed polymer is not necessary for our PPC method.

**4. Characterization of transferred vdW stack**

In Figs. 4(a–d), optical micrographs of the fabricated single- and bi-layer graphene and three- and four-layer h-BN on a thick h-BN/$SiO_2$/Si heterostructure are presented, respectively. Topographic image samples were measured by atomic force microscopy (AFM) and are respectively presented in Figs. 4(e–h). From both the optical micrograph and AFM image, it appears that in each case, the surface of the transferred graphene or h-BN is clean, without noticeable polymer residue. There are some bubbles presented in the transferred flakes. We infer that the number of bubbles can be reduced by further optimizing transfer conditions such as substrate temperature [10]. The thicknesses of the single- and bi-layer graphene and three- and four-layer h-BN on the thick h-BN in Figs. 4(e–h) were measured and the results are presented in Figs. 4(i–l), respectively. The thickness is determined to be about 0.5, 0.9, 1.2, and 1.5 nm, respectively. The thickness difference between the flakes is 0.3–0.4 nm and they are close to the single-layer thickness of both graphene and h-BN; thus, we think these prove the successful dry transfer of single- to few-layer-thick graphene and h-BN with a reasonable control of thickness.

Finally, we show the transport properties of graphene and h-BN devices fabricated with our proposed dry transfer method. By repeating the PPC-based dry transfer, encapsulated h-BN/graphene/h-BN and h-BN/graphene/few-layer hBN/graphene/h-BN structures were fabricated. By opting to use a top h-BN layer with a size smaller than the graphene layer, we allow the edge of the graphene to be not fully covered by the top h-BN layer. This enables us to construct a vdW heterostructure without using a one-dimensional edge-contact scheme [8]. This greatly simplifies the device fabrication as well as eliminates the requirement of fabricating of a global graphite gate with edge contact, which is difficult. We fabricated both top-contact (Fig. 5(a)) and edge-contact (Fig. 5(b)) h-BN/graphene/h-BN devices. In addition to this, a top-contact vertical-tunnel-junction device (Fig. 5(c)) was fabricated. Optical micrographs of the devices are presented in Figs. 5(d), 5(e), and 5(f), respectively. Note that top-contact device in Fig. 5(a) contains two graphene channels that are prepared by successive transfers: one uses a Si back-gate and the other uses a graphite back-gate. Magnetotransport measurements were performed at ~2.0 K. The longitudinal resistances were measured under the sweep of the carrier density $n$ by applying a gate voltage to the doped Si back-gate or graphite back-gate, and the results are presented in Figs. 5(g) and 5(h), respectively. The capacitance of the h-BN/SiO$_2$ dielectric was determined from the quantum Hall effect. The mobilities and charge inhomogeneity (extracted from the width of a charge neutrality point resistance peak) of the two top-contact h-BN/graphene/h-BN devices were calculated as ~260,000 cm$^2$/Vs and ~3×10$^{10}$ cm$^{-2}$ for back-gated device, ~140,000 cm$^2$/Vs and ~5×10$^{10}$ cm$^{-2}$ for graphite-gated device, respectively. These values are comparable to the mobility and the charge inhomogeneity of the edge contact device (Fig. 5(h)) of ~610,000 cm$^2$/Vs and ~5×10$^{10}$ cm$^{-2}$; thus, we think that quality of the device is not limited by the contact scheme. The extracted mobilities of the h-BN/graphene/h-BN devices approach that of modern h-BN/graphene/h-BN devices [7,8,10,37]; this indicates the cleanness of our transfer process. For the top-contacted vertical-tunnel-junction device shown in Fig. 5(f), we selected an h-BN tunnel barrier with a thickness of about 1.5 nm, as measured by AFM. Current–voltage (*I–V*) characteristics at about 2.0 K are presented in Fig. 5(i), showing a non-linear change in *I* with respect to *V*. The extracted junction resistance–area product *RA* is about 1×10$^{12}$ Ω·μm$^2$; this *RA* is comparable to that of the h-BN barrier with a similar thickness [38-40]. These results further prove that the PPC-based dry transfer method can be used to fabricate a multi-stack graphene/h-BN vdW heterostructure with clean interfaces.

## 5. Conclusions

We demonstrate the dry transfer fabrication of graphene as well as h-BN using the PPC-based method. The main advantages of this method are summarized in the following three points: 1) Because of the strong adhesion of PPC to graphene and h-BN at RT, appropriately large flakes of thin graphene and h-BN can be obtained via the mechanical exfoliation method. 2) Single-layer graphene, as well as thin h-BN, fabricated on a PPC/SiO$_2$/Si structure is visible under an optical microscope. 3) As the adhesion between the PPC and flakes significantly decreases at high temperatures, we are able to demonstrate the dry release transfer of graphene and h-BN on a h-BN/SiO$_2$/Si substrate at about 70 °C. Furthermore, we show that the fabricated graphene/h-BN devices exhibit excellent properties. So far, the PDMS-based dry transfer method has been widely used for fabricating TMD- and black phosphorus-based vdW heterostructures [5]. With the straightforward addition of a PPC sheet onto the PDMS, we demonstrate that graphene as well as few-layer h-BN can be also dry transferred. Our PPC-based dry transfer method is sufficiently simple and enables us to construct the heterostructure without using edge-contact scheme. Thus, our results further extend the capability of the dry release transfer method for constructing multilevel high-quality vdW heterostructures.

## 6. Methods

PPC (Sigma-Aldrich, CAS 25511-85-7) was dissolved into an anisole with a mass fraction of 15%. The PPC film was prepared by spin-coating the PPC solution onto a 290-nm-thick $SiO_2$/highly-doped Si substrate. The same substrate was used for the mechanical exfoliation of graphene and h-BN. A commercially available PDMS sheet (Gel-Pak PF-X4-17 mil.) was used. Graphene was mechanically exfoliated from a graphite crystal (NGS Naturgraphit GmbH) using adhesive tape (Nitto Denko, 32B) and was deposited on a PPC-coated $SiO_2$/Si substrate or $SiO_2$/Si substrate. Using the same method, thin h-BN flakes were exfoliated from its bulk crystal, which was synthesized by the high-pressure and high-temperature method and deposited onto the substrate. We chose thick h-BN with thicknesses in the range of 30–50 nm for the substrate and encapsulation for the graphene and thin h-BN. A commercial optical microscope (Keyence, VHX-5000) was used to observe fabricated flakes with a halogen lamp as a light source. To improve the adhesion of the PDMS, a plasma cleaner was operated with air as the plasma source (Harrick Plasma, PDC-32G). The thickness of the flakes was measured with an atomic force microscope (Hitachi High Technologies, SPM400). The device was spin-coated with a PMMA resist, and using electron beam (EB) lithography, electrode patterns were prepared. Au (~100 nm)/Cr (5 nm) contact electrodes were deposited by EB evaporation. For the edge-contact device, graphene and h-BN were etched using reactive ion etching with a $CF_4$-based gas. Immediately after plasma etching, the sample was loaded into the EB evaporation system and the electrodes were deposited. Transport properties were measured in a variable-temperature cryostat equipped with a superconducting magnet that can apply a magnetic field in the out-of-plane direction. After each dry transfer process, the vdW stacks were annealed in a tube furnace under the flow of Ar/3% $H_2$ gas.

## Acknowledgements

This work was supported by CREST, Japan Science and Technology Agency (JST) under Grant Number JPMJCR15F3, and by JSPS KAKENHI Grant Numbers JP25107001, JP25107003, JP25107004, JP26248061, JP15K21722, JP15K17433, and JP16H00982.

**Figure captions**

Figure 1

Schematics, calculated results, and optical micrographs of (a–e) graphene (Gr)/ poly(propylene) carbonate (PPC)/SiO$_2$/Si, (f–j) Gr/SiO$_2$/Si, (k–o) h-BN/PPC/SiO$_2$/Si, and (p–r) h-BN/SiO$_2$/Si structures. (a,f,k,p) Schematics of the device structure with refractive indexes $n_1$, $n_2$, $n_3$, $n_4$, and $n_5$ for Gr, PPC, SiO$_2$, Si, and h-BN, respectively. The thicknesses $d_1$, $d_2$, $d_3$, $d_4$, and $d_5$ are also depicted for Gr, PPC, SiO$_2$, Si, h-BN, respectively. (b,g,l,q) Calculated optical contrast as a function of wavelength and thickness for (b,l) PPC and (g,q) SiO$_2$. (c,h,m,r) Representative calculated optical contrast for single-layer (1L) and bi-layer (2L) (c,h) Gr and (m,r) h-BN. (d,e,i,j,n,o) Optical micrographs of (d,i) single-layer Gr, (e,j) bi-layer Gr, and (n,o) three-layer (3L) and four-layer (4L) h-BN.

Figure 2

(a–g) Schematics of dry transfer fabrication process. (a) Preparation of few-layer Gr on PPC/SiO$_2$/Si substrate. An optical micrograph of Gr/PPC/290-nm-thick SiO$_2$/Si is also shown. Separately, a piece of polydimethylpolysiloxane (PDMS) is prepared on the glass slide. (b) Preparation of tape window surrounding graphene area, with its optical micrograph shown in the top panel. Subsequently, this tape window, as well as the Gr/PPC sheet, is transferred onto the PDMS. An optical micrograph of PDMS on the glass slide is shown in the bottom panel. (c) Prepared Gr/PPC/PDMS structure on glass slide; the optical micrograph is also shown. (d) h-BN flake with thickness of about 30 nm prepared on SiO$_2$/Si substrate by mechanical exfoliation technique; the optical micrograph is also shown. (e) Adjusting the relative positions of Gr and h-BN flakes under an optical micrograph and making gentle contact at room temperature. (f) Heating the stage to 70 °C while graphene and h-BN are in contact. (g) Gentle separation of glass slide from SiO$_2$/Si substrate; the optical micrograph of the fabricated Gr/h-BN structure on the SiO$_2$/Si substrate is also shown.

Figure 3

(a) Storage modulus as a function of temperature for PPC and PDMS polymers. The data have been obtained from reference [30] and [34]. (b,c) Optical micrographs of Gr and graphite flakes on 900-nm-thick PPC/290-nm-thick SiO$_2$/Si, respectively. (d,e) Optical micrographs of transferred Gr and graphite flakes on 290-nm-thick SiO$_2$/Si, respectively.

Figure 4

(a–l) Optical micrograph, atomic force microscopy (AFM) topographic image, and AFM height profiles of (a,e,i) single-layer Gr on thick h-BN, (b,f,j) bi-layer Gr on thick h-BN, (c,g,k) three-layer h-BN on thick h-BN, and (d,h,l) four-layer h-BN on thick h-BN, respectively. The locations of the displayed AFM height profiles (i,j,k,l) are indicated by white lines in panels (e,f,g,h), respectively.

Figure 5

(a, b, c) Schematics of the device structure of (a) top-contacted h-BN/Gr/h-BN sample, (b) edge-contacted h-BN/Gr/h-BN sample, and (c) h-BN/Gr/few-layer h-BN/Gr/h-BN vertical-tunnel-junction sample. (d, e) Photographs of (d) top-contact and (e) edge-contact h-BN/Gr/h-BN devices. Dashed lines outline the location of graphene. (g,h) Resistance data as a function of carrier density of Gr for the devices shown in (d) and (e), respectively; conductivity data is shown in the insets. In panel (g), data shown in red and blue indicates the resistance value of left and right graphene in panel (d), respectively. (f) Photograph of h-BN/Gr/few-layer h-BN/Gr/h-BN vertical-tunnel-junction device. Solid and dashed red lines indicate the bottom and top graphene, respectively. The white line indicates the thin h-BN tunnel barrier. The thickness of few-layer h-BN is determined via AFM to be about 1.5 nm. (i) Current–voltage characteristics of the device shown in (c) at 3 K.

Figure 1

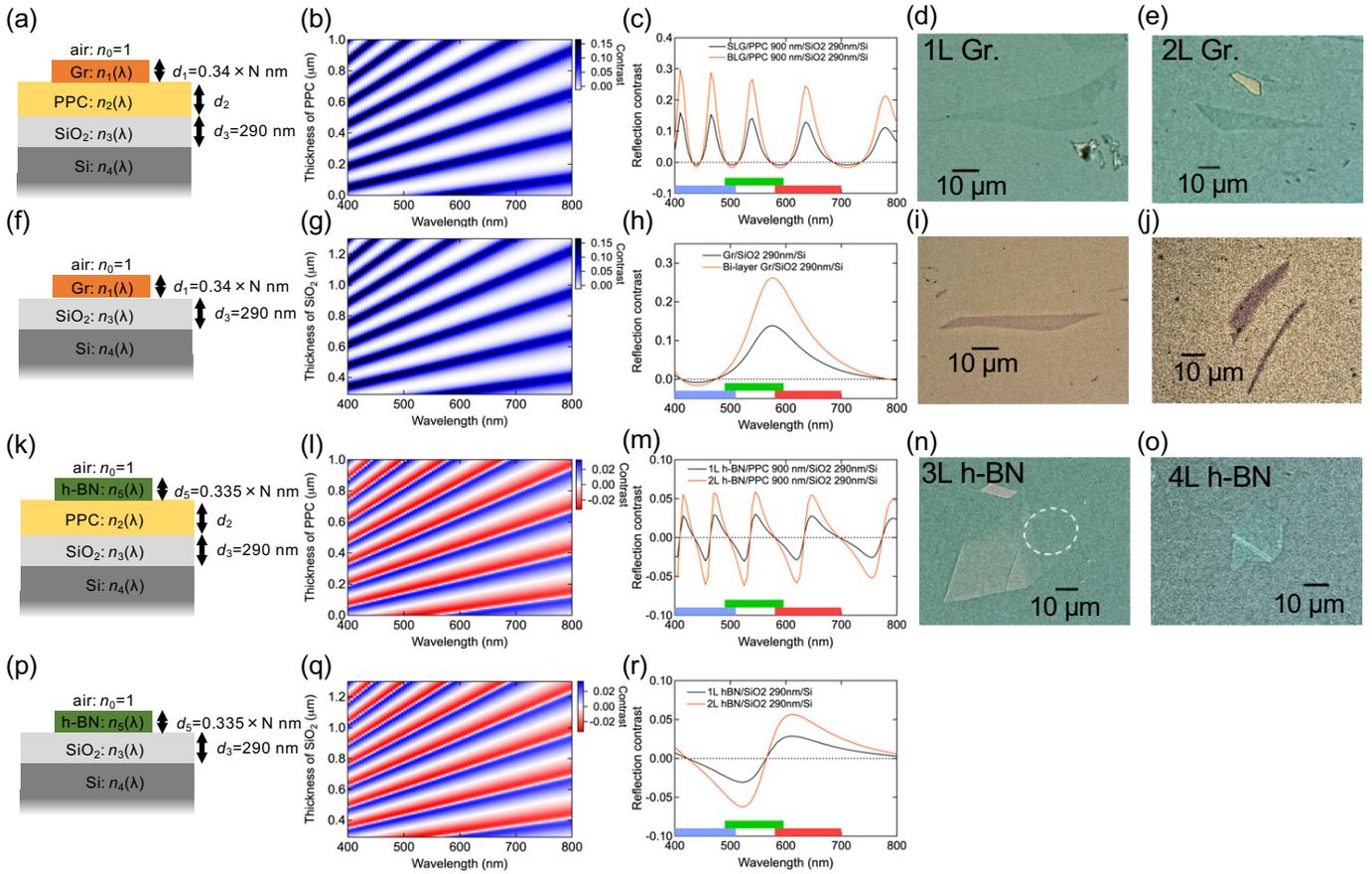

Figure 2

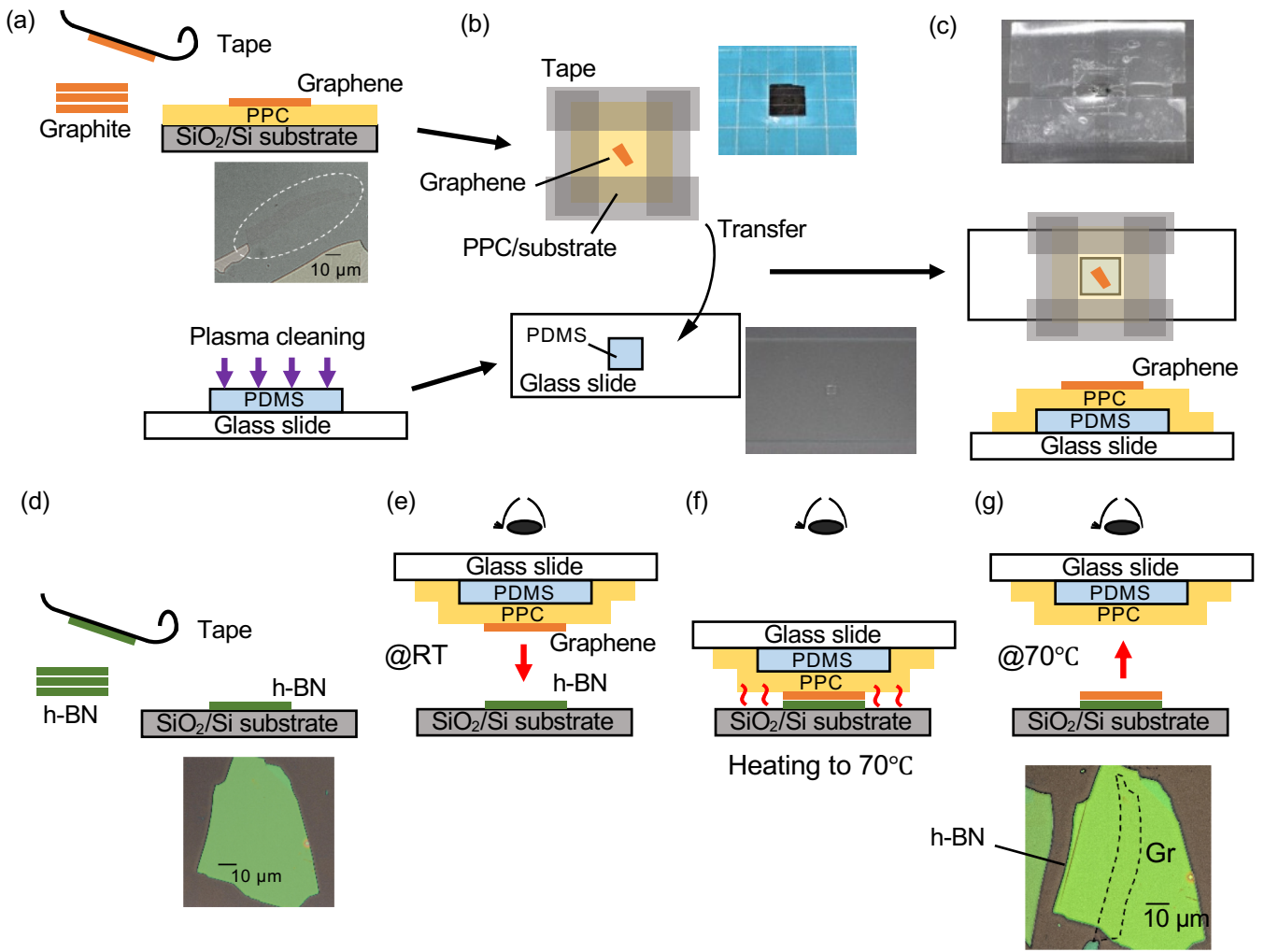

Figure 3

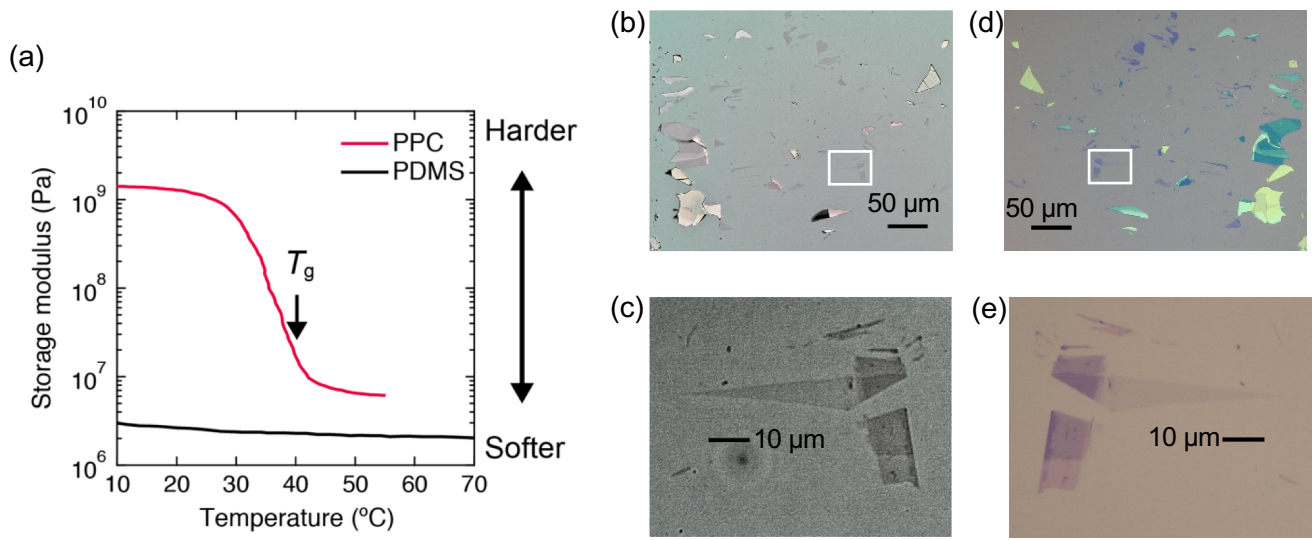

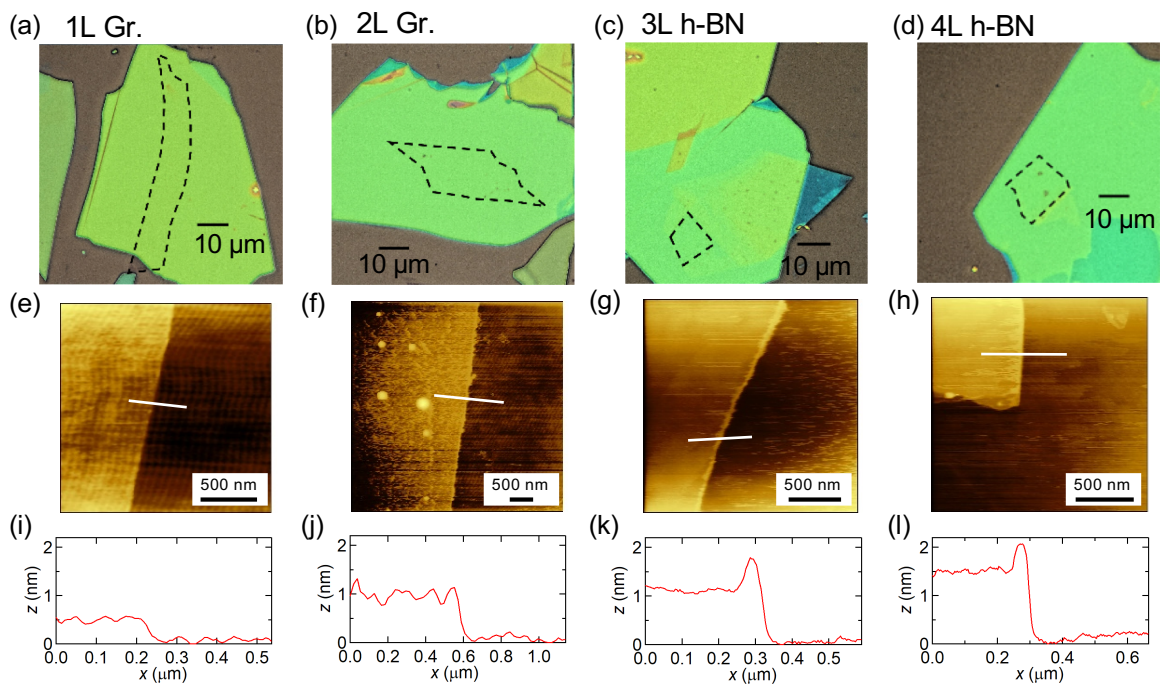

Figure 5

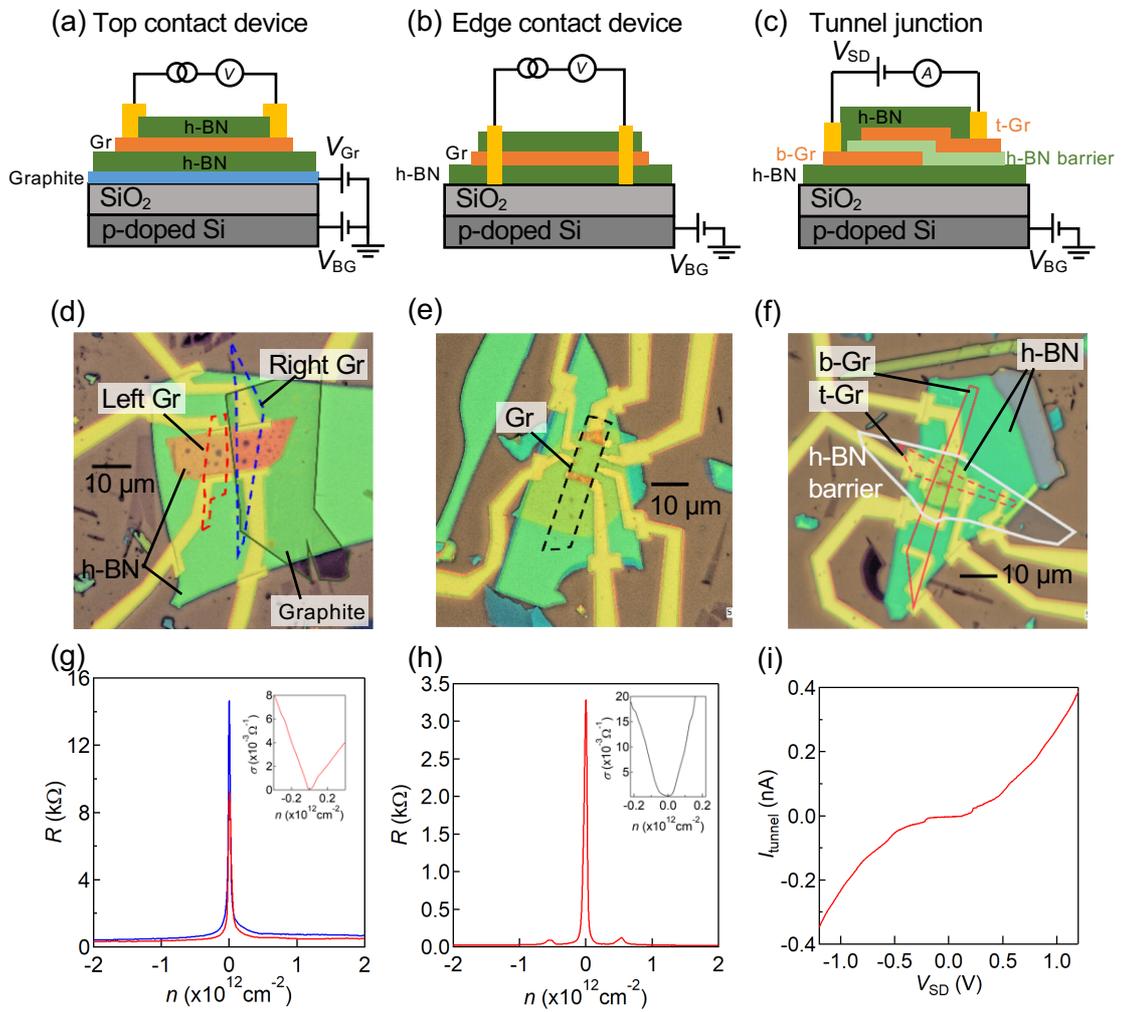